\documentclass[conference,onecolumn,12pt]{ieeetran}

\usepackage{mathrsfs,amsmath}
\usepackage{amsthm,amsfonts}
\usepackage{graphicx}

\newcommand{\bI}{\mathbf{I}}
\newcommand{\bM}{\mathbf{M}}
\newcommand{\bP}{\mathbf{P}}
\newcommand{\bQ}{\mathbf{Q}}
\newcommand{\bU}{\mathbf{U}}
\newcommand{\bV}{\mathbf{V}}
\newcommand{\bW}{\mathbf{W}}
\newcommand{\bX}{\mathbf{X}}
\newcommand{\bY}{\mathbf{Y}}
\newcommand{\bZ}{\mathbf{Z}}
\newcommand{\bp}{\mathbf{p}}
\newcommand{\bu}{\mathbf{u}}
\newcommand{\bv}{\mathbf{v}}

\newcommand{\bx}{\mathbf{x}}
\newcommand{\by}{\mathbf{y}}
\newcommand{\bz}{\mathbf{z}}
\newcommand{\bpi}{\boldsymbol{\Pi}}

\newtheorem{theorem}{Theorem}

\newtheorem{prop}[theorem]{Proposition}

\renewcommand{\baselinestretch}{1.3}

\title{Cooperative Repair of Multiple Node Failures in Distributed Storage Systems}

\author{
Kenneth W. Shum \\ Institute of Network Coding \\ The Chinese University of Hong Kong
\and Junyu Chen \\ Department of Information Engineering \\ The Chinese University of Hong Kong
}

\begin{document}

\maketitle

\begin{abstract}
Cooperative regenerating codes are designed for repairing multiple node failures in distributed storage systems.  In contrast to the original repair model of regenerating codes, which are for the repair of single node failure, data exchange among the new nodes is enabled. It is known that further reduction in repair bandwidth is possible with cooperative repair. Currently in the literature, we have an explicit construction of exact-repair  cooperative code achieving all parameters corresponding to the minimum-bandwidth point. We give a slightly generalized and more flexible version of this cooperative regenerating code in this paper. For minimum-storage regeneration with cooperation, we present an explicit code construction which can jointly repair any number of systematic storage nodes.
\end{abstract}


\section{Introduction}
In a distributed storage system, a data file is distributed to a number of storage devices that are connected through a network. The data is encoded in such a way that, if some of the storage devices are disconnected from the network temporarily, or break down permanently, the content of the file can be recovered from the remaining available nodes.  A simple encoding strategy is to replicate the data three times and store the replicas in three different places. This encoding method can tolerate a single failure out of three storage nodes, and is employed in large-scale cloud storage systems such as Google File System~\cite{GFS}. The major drawback of the triplication method is that the storage efficiency is fairly low. The amount of back-up data is two times that of the useful data.  As the amount of data stored in cloud storage systems is increasing in an accelerating speed, switching to encoding methods with higher storage efficiency is inevitable.

The Reed-Solomon (RS) code \cite{RS60} is a natural choice for the construction of  high-rate encoding schemes. The RS code is not only optimal, in the sense of being maximal-distance separable, it also has efficient decoding algorithms (see e.g. \cite{Roth}). Indeed, Facebook's storage infrastructure is currently employing a high-rate RS code with data rate 10/14. This means that four parity-check symbols are appended to every ten information symbols.  Nevertheless, not all data in Facebook's clusters is currently protected by RS code. This is because the traditional decoding algorithms for RS code do not take the network resources into account. Suppose that the 14 encoded symbols are stored in different disks. If one of the disks fails, then a traditional decoding algorithm needs to download 10 symbols from other storage nodes in order to repair the failed one. The amount of data traffic for repairing a single storage node is 10 times the amount of data to be repaired. In a large-scale distributed storage system, disk failures occur almost everyday~\cite{XORelephant}. The overhead traffic for repair would be prohibitive if all data were encoded by RS code.

In view of the repair problem, the amount of data traffic for the purpose of repair is an important evaluation metric for distributed storage systems. It is coined as the {\em repair bandwidth} by Dimakis {\em et al.} in \cite{DGWR07}. An erasure-correcting code with the aim of minimizing the repair bandwidth is called a {\em regenerating code}. Upon the failure of a storage node, we need to replace it by a new node, and the content of the new node is recovered by contacting $d$ other surviving nodes. The parameter $d$ is sometime called the {\em repair degree}, and the contacted nodes are called the {\em helper nodes} or simply the {\em helpers}. The repair bandwidth is measured by counting the  number of data symbols transmitted from the helpers to the new node.
If the data file can be reconstructed from any $k$ out of $n$ storage nodes, i.e., if any $n-k$ disk failures can be recovered, then we say that the {\em $(n,k)$-reconstruction property} is satisfied.  The design objective is to construct regenerating codes for $n$ storage nodes, satisfying the $(n,k)$-reconstruction, and minimizing the repair bandwidth, for a given set of code parameters $n$, $k$ and $d$.

We note that the requirement of $(n,k)$-reconstruction property is more relaxed than the condition of being maximal-distance separable (MDS). A regenerating code is an MDS erasure code only if the number of symbols contained in any $k$ nodes is exactly equal to the number of symbols in the data file. In a general regenerating code, the total number of coded symbols in any $k$ nodes may be larger than the total number of symbols in a data file.

There are two main categories of regenerating codes. The first one is called {\em exact-repair} regenerating codes, and the second one is called {\em functional-repair} regenerating codes. In the first category of exact-repair regenerating codes, the content of the new node is the same as in the old one.  In functional-repair regenerating codes, the content of the new node may change after a node repair, but the $(n,k)$-reconstruction property is preserved. For functional-repair regenerating code, a fundamental tradeoff between repair bandwidth and storage per node is obtained in~\cite{DGWR07}. This is done by drawing a connection to the theory of network coding. Following the notations in~\cite{DGWR07}, we denote the storage per node by $\alpha$ and the amount of data downloaded from a surviving node by $\beta$. The repair bandwidth is thus equal to $\gamma=d\beta $. A pair $(\alpha, d \beta )$ is said to be {\em feasible} if there is a regenerating code with storage $\alpha$ and repair bandwidth $d \beta $. It is proved in~\cite{DGWR07} that, for regenerating codes functionally repairing one failed node at a time,  $(\alpha,d \beta)$ is feasible if and only if the file size, denoted by $B$, satisfies the following inequality,
\begin{equation}
B \leq \sum_{i=0}^{k-1} \min\{\alpha, (d-i)\beta\}.
\label{eq:tradeoff}
\end{equation}
If we fix the file size $B$, the inequality in \eqref{eq:tradeoff} induces a tradeoff between storage and repair bandwidth.

There are two extreme points on the tradeoff curve. Among all the feasible pairs $(\alpha,d\beta)$ with minimum storage $\alpha$, the one with the smallest repair bandwidth is called the {\em minimum-storage regenerating} (MSR) point,
\begin{equation}
(\alpha_{\text{MSR}}, \gamma_{\text{MSR}}) = \Big(\frac{B}{k}, \frac{dB}{k(d+1-k)} \Big).
\label{eq:MSR}
\end{equation}
 On the other hand,  among all the feasible pairs $(\alpha,d \beta)$ with minimum bandwidth $d \beta$, the one with the smallest storage is called the {\em minimum-bandwidth regenerating} (MBR) point,
\begin{equation}
(\alpha_{\text{MBR}} , \gamma_{\text{MBR}}) = \Big(\frac{2dB}{k(2d+1-k)},\frac{2dB}{k(2d+1-k)} \Big).
\label{eq:MBR}
\end{equation}

Existence of linear functional-repair regenerating codes achieving all points on the tradeoff curve is shown in~\cite{Wu10}. Explicit construction of exact-repair regenerating codes, called the product-matrix framework, achieving all code parameters corresponding to the MBR point is given in \cite{productmatrix}. Explicit construction of regenerating codes for the MSR point is more difficult.
At the time of writing, we do not have constructions of exact-repair regenerating codes covering all parameters pertaining to the MSR point. Due to space limitation, we are not able to comprehensively review the literature on exact-repair MSR codes, but we mention below some constructions which are of direct relevance to the results in this paper.

The MISER code (which stands for MDS, Interference-aligning, Systematic Exact-Regenerating code) is an explicit exact-repair regenerating code at the MSR point~\cite{MISER} \cite{MISER_journal}. The code parameters are $d=n-1\geq 2k-1$. It is shown in \cite{MISER} and \cite{MISER_journal} that every systematic node, which contains uncoded data, can be repaired with storage and repair bandwidth attaining the MSR point in~\eqref{eq:MSR}. This result is extended in \cite{SR11}, which shows that, with the same code structure, every parity-check node can also be repaired with repair bandwidth meeting the MSR point. The product-matrix framework in~\cite{productmatrix} also gives a family of MSR codes with parameters $d \geq 2k-2$. All of the MSR codes mentioned above have code rate no more than $1/2$. For high-rate exact-repair MSR code, we refer the readers to three recent papers~\cite{SAK15}, \cite{RKV16} and \cite{YB16}, and the references contained therein.

We remark that the interior points on the tradeoff curve between storage and repair bandwidth for functional-repair regenerating codes are in general not achievable by exact-repair regenerating codes (see e.g. \cite{SRKR12} and~\cite{Tian433}).

All of the regenerating codes mentioned in the previous paragraphs are for the repair of a single node failure. In large-scale distributed storage system, it is not uncommon to encounter multiple node failures, due to various reasons. Firstly, the events of nodes failure may be correlated, because of power outage or aging. Secondly, we may not detect a node failure immediately when it happens. A scrubbing process is carried out periodically by the maintenance system, to scan the hard disks one by one and see whether there is any unrecoverable error. As the volume of the whole storage system increases, it will take a longer time to run the scrubbing process and hence the integrity of the disks will be checked less frequently. A disk error may remain dormant and undetected for a long period of time. If more than one errors occur during this period,  we will detect multiple disk errors during the scrubbing process. Lastly, in some commercial storage systems such as TotalRecall~\cite{Totalrecall}, the repair of a failed node is deliberately deferred.
During the period when some storage nodes are not available, degraded read is enabled by decoding the missing data in real time. A repair procedure is triggered after the number of failed nodes reaches a predetermined threshold. This mode of repair reduces the overhead of performing maintenance operations, and is called {\em lazy repair}.

A naive method for correcting multiple node failures is to repair the failed nodes one by one, using methods designed for repairing single node failure.  A collaborative recovery methodology for repairing multiple failed nodes jointly is suggested in~\cite{HXWZL10} and~\cite{WXHO10}. The repair procedure is divided into two phases. In the first phase, the new nodes download some repair data from some surviving nodes, and
in the second phase, the new nodes exchange data among themselves. The enabling of data exchange is the distinctive feature. We will call this the {\em cooperative} or {\em collaborative} repair model.

\begin{figure}
 \centering
 \includegraphics[width=3.5in]{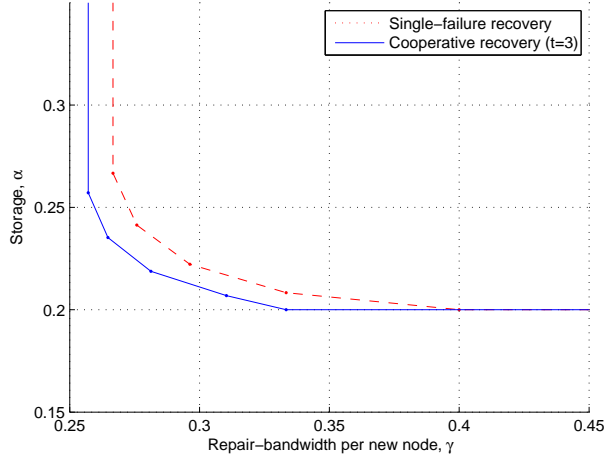}
 \caption{Tradeoff between storage and repair bandwidth for regenerating codes with parameters $d=8$, $k=5$, $B=1$, and $n\geq 11$. The dashed line is for regenerating code correcting single failure. The solid line is for cooperative regenerating code recovering $t=3$ failed nodes.}
 \label{figure}
\end{figure}

The minimum-storage regime for collaborative repair is considered in~\cite{HXWZL10} and~\cite{WXHO10}. It is shown that further reduction in repair bandwidth is possible if data exchange among the new nodes is allowed. Optimal function-repair minimum-storage regenerating codes are also presented in~\cite{WXHO10}. The results are extended by LeScouarnec  {\em et al.} to the opposite extreme point with minimum repair bandwidth in \cite{LeScouarnec} and \cite{KSS}.  The storage and repair bandwidth per new node on the minimum-storage collaborative regenerating (MSCR) point are denoted by $\alpha_{\text{MSCR}}$ and $\gamma_{\text{MSCR}}$, respectively, while the storage and repair bandwidth per new node on the  minimum-bandwidth collaborative regenerating (MBCR) point are denoted by $\alpha_{\text{MBCR}}$ and $\gamma_{\text{MBCR}}$, respectively.
The MSCR and MBCR points for functional repair are
\begin{align}
(\alpha_{\text{MSCR}}, \gamma_{\text{MSCR}}) &=\Big( \frac{B}{k}, \frac{B(d+t-1)}{k(d+t-k)} \Big),
\label{eq:MSCR} \\
(\alpha_{\text{MBCR}},  \gamma_{\text{MBCR}}) &=  \frac{B(2d+t-1)}{k(2d+t-k)}(1,1).
\label{eq:MBCR}
\end{align}
We note that when $t=1$, the operating points in \eqref{eq:MSCR} and \eqref{eq:MBCR} reduce to the ones in \eqref{eq:MSR} and \eqref{eq:MBR}.

The vertices on the tradeoff curve between storage and repair bandwidth for collaborative repair are characterized in~\cite{ShumHu13}. It is shown in \cite{ShumHu13} that for all points on the cooperative functional-repair tradeoff curve can be attained by linear regenerating codes over a finite field.
A numerical example of tradeoff curves for single-loss regenerating code and cooperative regenerating code is shown in Figure~\ref{figure}. We see that cooperative repair requires less  repair bandwidth in compare to single-failure repair.

Explicit exact-repair codes for the MBCR point for all legitimate parameters  were presented by Wang and Zhang in~\cite{WangZhang}. The construction in \cite{WangZhang} subsumes earlier constructions in \cite{ShumHu11b} and~\cite{Jiekak}.
In contrast, there are not so many explicit construction for MSCR code. The parameters of existing explicit constructions are summarized in Table~\ref{table:CRC}. A construction of exact repair for $k=t=2$ and $d = n-2$ is given in~\cite{LeScouarnec12}. This is extended to an MSCR code with $k\geq 2$ and $t=2$ in~\cite{ChenShum}. Indeed, a connection between MSCR codes which can repair $t=2$ node failures and non-cooperative MSR code is made in~\cite{LL14}. Using this connection, the authors in~\cite{LL14} are able to construct MSCR code with $t=2$ from existing MSR codes. However, there is no explicit construction for exact-repair MSCR code of any $t\geq 3$ failed nodes at the time of writing.

Practical implementations of distributed storage systems which can correct multiple node failures can be found in \cite{LL15} to \nocite{HDXQ15}\nocite{LLL15}\cite{PPR}.

\begin{table}
\caption{Parameters of explicit constructions of collaborative regenerating codes.}
\label{table:CRC}
\begin{center}
\begin{tabular}{|c|c|c|} \hline
Type & Code Parameters & Ref. \\ \hline \hline
MBCR & $n\geq d+t$, $d\geq k$, $t\geq 1$ & \cite{WangZhang} \\
\hline
MBCR & $n=d+t$, $d= k$, $t\geq 1$  & \cite{ShumHu11b} \\
\hline
MBCR & $n=d+t$, $d\geq k$, $t\geq 1$  & \cite{Jiekak} \\
\hline
MSCR & $n = d+2$, $k=t=2$ & \cite{LeScouarnec12} \\ \hline
MSCR & $n = 2k$, $d=n-2$, $k\geq 2$, $t=2$ & \cite{ChenShum} \\ \hline
MSCR & $n = 2k$, $d=n-t$, $k\geq 2$, $k\geq t\geq 2$ & \cite{ChenShum} \\
 & (repair of systematic nodes only)&\\ \hline
\end{tabular}
\end{center}
\end{table}

The rest of this paper is organized as follows. In Section~\ref{sec:model}, we formally define linear regenerating codes for distributed storage systems with collaborative repair. In Section~\ref{sec:MBCR}, we give a slight generalization of the cooperative regenerating  codes in \cite{WangZhang}. The generalized version also achieves all code parameters of the MBCR point, but the building blocks of the construction  only need to satisfy a more relaxed condition. In Section~\ref{sec:MSCR}, we give a simplified description of the repair method in~\cite{ChenShum}, and illustrate how to repair two or more systematic nodes collaboratively in the MISER code. Some concluding remarks are listed in Section~\ref{sec:conclusion}.

\section{A Collaborative Repair Model for Linear Regenerating Code}
\label{sec:model}

We will use the following notations in this paper:
\smallskip

\noindent $B$: file size.

\noindent $n$: the total number of storage nodes.

\noindent $k$: the number of storage nodes from which a data collector can decode the original file.

\noindent $d$: The number surviving nodes contacted by a new node.

\noindent $t$: the number of new nodes we want to  repair collaboratively.

\noindent $\alpha$: the amount of data stored in a node.

\noindent $\beta_1$: the amount of data downloaded from a helper node to a new node during the first phase of repair.

\noindent $\beta_2$: the amount of data exchanged between two new node during the second phase of repair.

\noindent $\gamma$: the repair bandwidth per new node.

\noindent $\mathbb{F}_q$: finite field of size $q$, where $q$ is a prime power.

\smallskip

We describe in this section a mathematical formulation of linear collaborative exact repair. For the problem formulation for the non-linear case, we refer the readers to~\cite{ShumHu13}.

A data file consists of $B$ symbols.
We let $M$ be the vector space $\mathbb{F}_q^B$. We regard a data file as a vector in $M$, and call it the source vector~$\mathbf{m}$.

The source vector $\mathbf{m}$ is mapped to $n \alpha$ finite field symbols, and each node stores $\alpha$ of them. The mapping from the source vector $\mathbf{m}$ to an encoded symbol is a linear functional on $M$.
Following the terminology of network coding, we will call these linear mappings the {\em encoding vectors} associated to the encoded symbols. Formally, a linear functional is an object in the dual space of $M$, $L(M, \mathbb{F}_q)$, which consists of all linear transformations from $M$ to $\mathbb{F}_q$. More precisely, an encoding vector should be called an encoding {\em co-vector} instead, but we will be a little bit sloppy on this point and simply use the term ``vector''.

The content of a storage node can be described by a subspace of $L(M,\mathbb{F}_q)$, spanned by the encoding vectors of the encoded symbols stored in this node. For $i=1,2,\ldots, n$, we let $W_i$ denote the subspace of $L(M,\mathbb{F}_q)$ pertaining to node $i$. The dimension of $W_i$ is no more than $\alpha$,
$$
dim(W_i) \leq \alpha
$$
for all $i$.

We want to distribute the data file to the $n$ storage nodes in such a way that any $k$ of them are sufficient in reconstructing the source vector $\mathbf{m}$. The $(n,k)$-reconstruction property requires that the $k\alpha$ encoding vectors in any $k$ storage nodes span the dual space  $L(M,\mathbb{F}_q)$, hence it is required that
$$
\bigoplus_{i\in\mathcal{K}} W_i = L(M,\mathbb{F}_q),
$$
for any $k$-subset $\mathcal{K}$ of $\{1,2,\ldots, n\}$. Here $\bigoplus_i W_i$ denotes the sum space of $W_i$'s. It will be a direct sum if the regenerating code is MDS.

Suppose that the storage nodes with indices $i_1$, $i_2,\ldots, i_t$ fail, and we need to replace them by $t$ new nodes. For $s=1,2,\ldots, t$, new node $s$ contacts $d$ available nodes, and download $\beta_1$ symbols from each of them. The storage nodes which participate in the repair process are called the {\em helpers}. Different new nodes may download repair data from different sets of helpers.  Let $\mathcal{H}_s$ be the index set of the $d$ helpers contacted by new node $s$. Thus, we have
$$\mathcal{H}_s \subseteq \{1,2,\ldots, n\} \setminus \{i_1, i_2,\ldots, i_t\}
$$
and $|\mathcal{H}_s|=d$ for all $s$. The downloaded symbols are linear combination of the symbols kept by the helpers.  The encoding vector of a symbol downloaded from node $j$ is thus contained in~$W_j$. For $s=1,2,\ldots, t$, let $U_s$ be the subspace of $L(M,\mathbb{F}_q)$ spanned by the $d \beta_1$ encoding vectors of the symbols sent to new node $s$. We have
$$
dim(U_s \cap W_j) \leq \beta_1,
$$
for all $s=1,2,\ldots, t$ and $j\in \mathcal{H}_s$.

In the second phase of the repair, new node $s$ computes and sends $\beta_2$ finite field symbols to new node $s'$, for $s,s'\in\{1,2,\ldots, t\}$ and $s\neq s'$. The computed symbols are linear combinations of the symbols which are already received by new node $s$ in the first phase of repair. Let $V_{s \rightarrow s'}$ be the subspace of $L(M,\mathbb{F}_q)$ spanned by the encoding vectors of the symbols sent from node $s$ to node $s'$ during the second phase. We have
$$
 V_{s \rightarrow s'} \subseteq U_s, \text{ and } dim(V_{s \rightarrow s'}) \leq \beta_2.
$$
For $s'=1,2,\ldots, t$, new node $s'$ should be able to recover the content of the failed node $i_{s'}$. In terms of the subspaces, it is required that
$$
W_{i_{s'}} \subseteq U_{s'} \oplus \bigoplus_{s \in \{1,2,\ldots, t\}\setminus\{s'\}} V_{s\rightarrow s'}.
$$
The repair bandwidth per new node is equal to
$$
\gamma  = d\beta_1 + (t-1)\beta_2.
$$
Any linear code satisfying the above requirements is called a {\em cooperative regenerating code} or {\em collaborative regenerating code}.

\section{Cooperative Regenerating Codes with Minimum Repair Bandwidth}
\label{sec:MBCR}

In this section we give a slight generalization of the construction of minimum-bandwidth cooperative regenerating codes in~\cite{WangZhang}. The number of failed nodes, $t$,  to be repaired jointly can be any positive integer. The code parameters which can be supported by the construction to be described below is the same as those in~\cite{WangZhang}, i.e.,  $n$, $k$ and $d$ satisfy
$$
 n-t \geq d  \geq k.
$$
The file size $B$ of the regenerating code is
$$
B = k(2d+t-k),
$$
and each storage node stores $2d+t-1$ symbols.
In contrast to the polynomial approach in~\cite{WangZhang}, the construction below depends on the manipulation of a bilinear form (to be defined in \eqref{eq:bilinear_form}).


\medskip

\noindent {\bf Encoding.} We need a $d\times n$ matrix $\bU$ and a $(d+t)\times n$ matrix $\bV$ for the encoding. Partition $\bU$ and $\bV$ as
$$
\bU = \begin{bmatrix}
\bU_1\\ \hline
\bU_2
\end{bmatrix}, \ \bV=
\begin{bmatrix}
\bV_1\\ \hline
\bV_2
\end{bmatrix},
$$
where $\bU_1$ and $\bV_1$ are submatrices of  size $k \times n$. We will choose the matrices $\bU$ and $\bV$  such that the following conditions are satisfied:

\begin{enumerate}
\item any $d\times d$ submatrix of $\bU$ is nonsingular;

\item any $(d+t) \times (d+t)$ submatrix of $\bV$ is non-singular;

\item any $k\times k$ submatrix of $\bU_1$ is nonsingular;

\item any $k\times k$ submatrix of $\bV_1$ is nonsingular.
\end{enumerate}

We can obtain matrices $\bU$ and $\bV$ by Vandermonde matrix or Cauchy matrix. If we use Vandermonde matrix, we can set the $i$-th column of $\bU$ to
$$
  \begin{bmatrix}
	1 & x_i & x_i^2 & & \ldots&  & x_i^{d-1}
	\end{bmatrix}^T,
$$
for $i=1,2,\ldots, n$.
If $x_1, x_2,\ldots, x_n$ are distinct elements in $\mathbb{F}_q$, then the resulting matrix $\bU$ satisfies the first and third conditions listed above. We can use Vandermonde matrix for the matrix $\bV$ similarly.
Existence of such matrices is guaranteed if the field size is larger than or equal to~$n$. Anyway, the correctness of the code construction only depends on the four conditions above.

For $i=1,2,\ldots, n$, we denote the $i$-th column of $\bU$ by $\bu_i$, and the $i$-th column of $\bV$ by $\bv_i$.

We arrange the source symbols
in a $d\times (d+t)$ partitioned matrix
$$
\mathbf{M} = \left[ \begin{array}{c|c}
\mathbf{A}& \mathbf{B} \\ \hline
\mathbf{C} & \mathbf{0} \\
\end{array} \right],
$$
where $\mathbf{A}$, $\mathbf{B}$ and $\mathbf{C}$ are sub-matrices of size $k\times k$, $k\times (d+t-k)$ and $(d-k)\times k$, respectively.  The total number of entries in the three sub-matrices is
\[
 k^2 + k(d+t-k) + (d-k)k  = k(2d+t-k) = B.
\]
We will call $\bM$ the {\em source matrix}.

The source matrix $\bM$ induces a bilinear form $\mathsf{B}$
defined by
\begin{equation}
\mathsf{B}(\bx, \by) := \bx^T \bM \by,
\label{eq:bilinear_form}
\end{equation}
for $\bx\in\mathbb{F}_q^d$ and $\by\in\mathbb{F}_q^{d+t}$.
We distribute the information to the storage nodes in such a way that, for $i=1,2,\ldots, n$,
node $i$ is able to compute the following two linear functions,
$$
\mathsf{B}(\cdot, \bv_i) \text{ and } \mathsf{B}(\bu_i, \cdot).
$$
The first one is a linear mapping from $\mathbb{F}_q^d$ to $\mathbb{F}_q$, and the second is from $\mathbb{F}_q^{d+t}$ to $\mathbb{F}_q$. Node $i$ can store the $d$ entries in the vector $\bM \bv_i$, and compute the first function $\mathsf{B}(\cdot, \bv_i)$ by taking the inner product of the input vector $\bx$ and $\bM \bv_i$,
$$
\mathsf{B}(\bx, \bv_i) = \bx^T ( \bM \bv_i).
$$
For the second function $\mathsf{B}(\bu_i, \cdot)$, node $i$ can store the $d+t$ entries in the vector $\bu_i^T \bM$, and compute $\mathsf{B}(\bu_i, \by)$ by
$$
\mathsf{B}(\bu_i, \by) = (\bu_i^T \bM ) \by.
$$
Since the components of $\bM \bv_i$ and $\bu_i^T \bM $ satisfy a simple linear equation,
\begin{equation}
\bu_i (\bM \bv_i) - (\bu_i^T \bM) \bv_i = 0,
\label{eq:missing}
\end{equation}
we only need to store $d+(d-t)-1$ finite field elements in node $i$, in order to implement the function $\mathsf{B}(\cdot, \bv_i)$ and $\mathsf{B}(\bu_i,\cdot)$. Hence, each storage node  is only required to store
$$
 \alpha = 2d+t-1
$$
finite field elements.

\medskip

\noindent {\bf Repair procedure.} Without loss of generality, suppose that nodes 1 to $t$ fail. For $i=1,2,\ldots, t$, the $i$-th new node downloads some repair data from a set of $d$ surviving nodes, which can be chosen arbitrarily. Let $\mathcal{H}_i$ be the index set of the $d$ surviving nodes contacted by node $i$. We have $\mathcal{H}_i \subseteq \{t+1,t+2,\ldots, n\}$ and $|\mathcal{H}_i|=d$ for all $i$. The helper with index $j\in \mathcal{H}_i$ computes two finite field elements
$$
\mathsf{B}(\bu_i,\bv_j) \text{ and } \mathsf{B}(\bu_j,\bv_i),
$$
and transmits them to new node $i$. In the first phase of repair, a total of $2dt$ symbols are transmitted from the helpers.

For $i=1,2,\ldots, t$, the $i$-th new node can recover $\bM \bv_i$ from the following $d$-dimensional vector with the $d$ components indexed by $\mathcal{H}_j$.
$$
( \bu_j^T \bM \bv_i)_{j\in\mathcal{H}_i} = [\bu_j^T]_{j\in\mathcal{H}_i} \cdot (\bM \bv_i),
$$
where  $[\bu_j^T]_{j\in\mathcal{H}_i}$ is the  $d\times d$ matrix obtained by stacking the row vectors $\bu_j^T$ for $j\in \mathcal{H}_i$. Since this matrix is nonsingular by construction, the $i$-th new node can obtain $\bM \bv_i$. At this point, the $i$-th new node is able to compute the function $\mathsf{B}(\cdot, \bv_i)$.

In the second phase of the repair procedure, node~$i$ calculates $\mathsf{B}(\bu_\ell, \bv_i)$, for $\ell\in \{1,2,\ldots, t\}\setminus \{i\}$, and sends the resulting finite field symbol to the $\ell$-th new node.
Furthermore, node $i$ can compute $\mathsf{B}(\bu_i,\bv_i)$, using the information already obtained from the first phase of repair. Node $i$ can now calculate $\bu_i^T\bM$ from
$$
\bu_i^T \bM \bv_s, \ \text{ for }s \in \mathcal{H}_i\cup \{1,2,\ldots, t\},
$$
using the property that the vectors $\bv_s$, for $s \in \mathcal{H}_i \cup  \{1,2,\ldots, t\}$, are linearly independent over $\mathbb{F}_q$. The repair of node $i$ is completed by storing $2d+t-1$ components in the vectors $\bu_i^T \bM$ and $\bM \bv_i$, which are necessary in computing $\mathsf{B}(\cdot, \bv_i)$ and $\mathsf{B}(\bu_i, \cdot)$.

We remark that the total number of transmitted symbols in the whole repair procedure is $2dt+t(t-1)$, and therefore the repair bandwidth per new node is
$$
\gamma = 2d+t-1.
$$

\medskip

\noindent {\bf File recovery.} Suppose that a data collector connects to nodes $i_1, i_2,\ldots, i_k$, with
$$
1 \leq i_1 <i_2 < \cdots < i_k \leq n.
$$
The data collector can download the vectors
$$
\bM \bv_{i_\ell} \text{ and } \bu^T_{i_\ell} \bM,
$$
for $\ell=1,2,\ldots, k$.
From the last $d-k$ of the components in $\bM \bv_{i_\ell}$, for $\ell=1,2,\ldots, k$, we can recover the $(d-k)\times k$ sub-matrix $\mathbf{C}$ in the source matrix $\bM$, because any $k\times k$ submatrix $\bV_1$ of $\bV$ is nonsingular by assumption. Similarly, from the last $d+t-k$ components in $\bu_{i_\ell}^T \bM$, we can recover the $(d+t-k)\times k$ sub-matrix $\mathbf{B}$, using the property that any $k\times k$ submatrix $\bU_1$ is nonsingular.
The remaining source symbols in $\mathbf{A}$ can be decoded either from the first $k$ components of vectors $\bM \bv_{i_\ell}$, or the first $k$ components of the vectors $\bu_{i_\ell}^T \bM$.

{\bf Example.}
We illustrate the construction by the following example with code parameters
$n=7$, $d=4$, $k=t=3$. The file size is
$B=k(2d+t-k) = 24$. In this example, we pick $\mathbb{F}_7$ as the underlying finite field.

The source matrix is partitioned as
$$\bM = \left[ \begin{array}{ccc|cccc}
a_{11} & a_{12} & a_{13} & b_{11} & b_{12} & b_{13} & b_{14} \\
a_{21} & a_{22} & a_{23} & b_{21} & b_{22} & b_{23} & b_{24} \\
a_{31} & a_{32} & a_{33} & b_{31} & b_{32} & b_{33} & b_{34} \\ \hline
c_{11} & c_{12} & c_{13} & 0 & 0 & 0 & 0 \\
\end{array} \right].
$$
The entries $a_{ij}$'s, $b_{ij}$'s and $c_{ij}$'s are the source symbols.
Let $\mathsf{B}(\bx,\by)$ be the bilinear form defined as in~\eqref{eq:bilinear_form}, mapping  a pair of vectors $(\bx,\by)$ in $\mathbb{F}_7^4 \times \mathbb{F}_7^7$ to an element in~$\mathbb{F}_7$.

Let $\bU$ be the $4\times 7$ Vandermonde matrix
\begin{align}
\bU &= \begin{bmatrix}
 1 & 1 & 1 & 1 & 1 & 1 & 1\\
 1 & 2 & 3 & 4 & 5 & 6 & 0\\
 1 & 4 & 2 & 2 & 4 & 1 & 0\\
 1 & 1 & 6 & 1 & 6 & 6 & 0
\end{bmatrix}
\label{eq:UU}
\end{align}
and for $i=1,2,\ldots, n$, let $\mathbf{u}_i = \begin{bmatrix} 1 & i & i^2 & i^3 \end{bmatrix}^T$ be the $i$-th column of $\bU$.
Let $\bV$ be the $7\times 7$ Vandermonde matrix
\begin{align}
\bV &= \begin{bmatrix}
 1 & 1 & 1 & 1 & 1 & 1 & 1\\
 1 & 2 & 3 & 4 & 5 & 6 & 0\\
 1 & 4 & 2 & 2 & 4 & 1 & 0\\
 1 & 1 & 6 & 1 & 6 & 6 & 0\\
 1 & 2 & 4 & 4 & 2 & 1 & 0\\
 1 & 4 & 5 & 2 & 3 & 6 & 0\\
 1 & 1 & 1 & 1 & 1 & 1 & 0
\end{bmatrix} \label{eq:VV}
\end{align}
and for $i=1,2,\ldots, n$, let $\mathbf{v}_i = \begin{bmatrix} 1 & i & i^2 & \ldots & i^6 \end{bmatrix}^T$ be the $i$-th column of $\bV$.
The $i$-th node needs to store enough information such that it can compute the functions
$$
\mathsf{B}(\cdot, \bv_i) \text{ and } \mathsf{B}(\bu_i,\cdot).
$$
For instance, node $i$ can store the last  3 components in vector $\bM \bv_i$, and all 7 components in $\bu_{i}^T \bM$,
\begin{align*}
z_{i1} &:=a_{21} + ia_{22}+i^2a_{23}+i^3b_{21}+i^4b_{22}+i^5b_{23}+i^6 b_{24} ,\\
z_{i2} &:=a_{31} + ia_{32}+i^2a_{33}+i^3b_{31}+i^4b_{32}+i^5b_{33}+i^6 b_{34} ,\\
z_{i3} &:=c_{11} + ic_{12}+i^2c_{13} ,\\
z_{i4} &:=a_{11} + ia_{21}+i^2a_{31}+i^3 c_{11}, \\
z_{i5} &:=a_{12} + ia_{22}+i^2a_{32}+i^3 c_{12}, \\
z_{i6} &:=a_{13} + ia_{23}+i^2a_{33}+i^3 c_{13}, \\
z_{i7} &:=b_{11} + ib_{21}+i^2b_{31},\\
z_{i8} &:=b_{12} + ib_{22}+i^2b_{32},\\
z_{i9} &:=b_{13} + ib_{23}+i^2b_{33}, \\
z_{i10} &:= b_{14} + ib_{24}+i^2b_{34},
\end{align*}
with all arithmetic performed modulo $7$. The missing entry of $\bM \bv_i$, namely, the first entry of $\bM \bv_i$,
\begin{align*}
&a_{11} + ia_{12}+i^2a_{13}+i^3b_{11}+i^4b_{12}+i^5b_{13}+i^6 b_{14} \\
&=
-iz_{i1}-i^2z_{i2}-i^3 z_{i3}  \\
 &\qquad + z_{i4}+ i z_{i5}+i^2 z_{i6}+i^3 z_{i7}+ i^4 z_{i8}+ i^5 z_{i9} + i^6 z_{i10}
\end{align*}
is a linear combination of $z_{i1},z_{i2},\ldots z_{i10}$. Each node only needs to store 10 finite field symbols $z_{i1},z_{i2},\ldots, z_{i10}$. The storage per node meets the bound
$$
\alpha_{\text{MBCR}} = \frac{B(2d+t-1)}{k(2d+t-k)} = 2d+t-1 = 10.
$$

We illustrate the repair procedure by going through the repair of nodes 5, 6 and 7. Suppose we lost the content of nodes 5, 6 and 7, and want to rebuild them by cooperative repair.
For $i=1,2, 3, 4$, and $j=5,6,7$, node $i$ computes $\mathsf{B}(\bu_i, \bv_j)$ and $\mathsf{B}(\bu_j, \bv_i)$ and sends them to the node $j$, in the first phase of repair. Node $j$ now have 8 symbols,
\begin{gather*}
\mathsf{B}(\bu_1, \bv_j),\
\mathsf{B}(\bu_2, \bv_j),\
\mathsf{B}(\bu_3, \bv_j),\
\mathsf{B}(\bu_4, \bv_j),\\
\mathsf{B}(\bu_j, \bv_1),\
\mathsf{B}(\bu_j, \bv_2),\
\mathsf{B}(\bu_j, \bv_3),\
\mathsf{B}(\bu_j, \bv_4).\
\end{gather*}
The first four of them can be put together and form a vector
$$
\begin{bmatrix}
\mathsf{B}(\bu_1,\bv_j) \\
\mathsf{B}(\bu_2,\bv_j) \\
\mathsf{B}(\bu_3,\bv_j) \\
\mathsf{B}(\bu_4,\bv_j)
\end{bmatrix}
= \begin{bmatrix}
\bu_1^T \\
\bu_2^T \\
\bu_3^T \\
\bu_4^T
\end{bmatrix}
\bM \bv_j.
$$
Because the first four columns of matrix $\bU$ in \eqref{eq:UU} are linearly independent over $\mathbb{F}_7$, for $j=5,6,7$, node $j$ can solve for $\bM \bv_j$ after the first phase of repair, and is able to calculate $\mathsf{B}(\bx,\bv_j)$ for any vector $\bx \in \mathbb{F}_7^4$.

The communications among nodes 5, 6 and 7 in the second phase of repair is as follows:

\noindent node 5 sends $\mathsf{B}(\bu_6, \bv_5)$ to node 6,

\noindent node 5 sends $\mathsf{B}(\bu_7, \bv_5)$ to node 7,

\noindent node 6 sends $\mathsf{B}(\bu_5, \bv_6)$ to node 5,

\noindent node 6 sends $\mathsf{B}(\bu_7, \bv_6)$ to node 7

\noindent  node 7 sends $\mathsf{B}(\bu_5, \bv_7)$ to node 5,

\noindent node 7 sends $\mathsf{B}(\bu_6, \bv_7)$ to node 6.

For $j=5,6,7$, node $j$ can obtain $\bu_j^T \bM$ from
$$
\bu_j^T \bM \begin{bmatrix}
\bv_1 & \bv_2& \bv_3 & \bv_4& \bv_5 & \bv_6 & \bv_7
\end{bmatrix}.
$$
In the first phase, we transmit $4\cdot3\cdot2=24$ symbols, and in the second phase we transmit 6 symbols. The number of transmitted symbol per new node is thus equal to 10, which is equal to the target repair bandwidth $\gamma =2d+t-1=10$.

To illustrate the $(n,k)$-reconstruction property, suppose that a data collector connects to nodes 1, 2 and~3. The data collector can download the following vectors
$$
\bu_1^T \bM, \ \bu_2^T \bM ,\ \bu_3^T \bM, \
\bM \bv_1,\ \bM \bv_2,\text{ and } \bM \bv_3.
$$
There are totally 33 symbols in these six vectors. They are not linearly independent as the original file only contains 24 independent symbols. We can decode the symbols in the data file by selecting 24 entries in the received vectors, and form a vector which can be written as the product of a $24 \times 24$ lower-block-triangular matrix and a 24-dimensional vector
$$
\begin{bmatrix}
\bV_3 & & \\
\mathbf{0} & \bV_3 &  \\
\mathbf{0} &\mathbf{0} & \bV_3 & \\
\mathbf{0} &\mathbf{0} &\mathbf{0} & \bV_3 & \\
\mathbf{0} &\mathbf{0} &\mathbf{0} &\mathbf{0} & \bV_3 & \\
\mathbf{D} &\mathbf{0}&\mathbf{0}&\mathbf{0}&\mathbf{0}& \bV_3 & \\
\mathbf{D} &\mathbf{0}&\mathbf{0}&\mathbf{0}&\mathbf{0}& \mathbf{0}&\bV_3 \\
\mathbf{D} &\mathbf{0}&\mathbf{0}&\mathbf{0}&\mathbf{0}&\mathbf{0}&\mathbf{0}& \bV_3 \\
\end{bmatrix}
\begin{bmatrix}
c_{11} \\c_{12} \\c_{13} \\\hline
b_{11} \\b_{21} \\b_{31} \\\hline
\vdots \\ \hline
b_{14} \\b_{24} \\b_{34} \\\hline
a_{11} \\a_{21} \\a_{31} \\\hline
\vdots
\end{bmatrix}
$$
with $\bV_3$ denoting a $3\times 3$ nonsingualr Vandermonde matrix, and $\mathbf{D}$ a diagonal matrix. The above matrix is invertible and we can obtain the source symbols in the data file.

\section{A Class of Minimum-Storage Cooperative Regenerating Codes}
\label{sec:MSCR}
In this section, we give a simplified description of the the minimum-storage cooperative regenerating code presented in~\cite{ChenShum}. The code parameters are
$$
n = 2k, \ d = n-t,\ k \geq t \geq 2.
$$
The first $k$ nodes are the systematic nodes, while the last $k$ nodes are the parity-check nodes. The coding structure of the cooperative regenerating codes to be described in this section is indeed the same as the MISER code~\cite{MISER},\cite{MISER_journal} and the regenerating code in~\cite{SR11}. Our objective is to show that, with this coding structure, we can repair the failure of any $t$ systematic nodes and any $t$ parity-check nodes, for any $t$ less than or equal to $k$, attaining the MSCR point defined in~\eqref{eq:MSCR}.

We need a nonsingular matrix $\bU$ and a super-regular matrix $\bP$, both of size $k\times k$. Recall that a matrix is said to be super-regular if every square submatrix is nonsingular. Cauchy matrix is an example of super-regular matrix, and we may let $\bP$ be a Cauchy matrix.

After the matrices $\bU$ and $\bP$ are fixed, we let $\bQ$ be the inverse of $\bP$ and $\bV$ be the matrix $\bV := \bU \bP$. It can be shown that the matrix $\bV$ is non-singular and $\bQ$ is super-regular.
We have the following relationship among these matrices
$$
 \bV = \bU \bP \text{ and } \bU = \bV \bQ.
$$
Let $p_{ij}$ be $(i,j)$-entry of $\bP$, for $i,j \in \{1,2,\ldots, k\}$, and $q_{ij}$ be the $(i,j)$-entry of $\bQ$.

For $i=1,2,\ldots, k$, let $\bu_i$ denote the $i$-th column of $\bU$, and $\bv_i$ the $i$-th column of $\bV$. The columns of $\bU$ and the columns of $\bV$ will be regarded as two bases of vector space $\mathbb{F}_q^k$. Let $\hat{\bu}_1, \hat{\bu}_2,\ldots, \hat{\bu}_k$  be the dual basis of $\bu_i$'s, and let $\hat{\bv}_1, \hat{\bv}_2,\ldots, \hat{\bv}_k$  be the dual basis of $\bv_i$'s. The dual bases satisfy the following defining property
$$\hat{\bu}_i^T \bu_j = \delta_{ij}, \text{ and } \hat{\bv}_i^T \bv_j = \delta_{ij},
$$
where $\delta_{ij}$ is the Kronecker delta function.

The last ingredient of the construction is a $2\times 2$ super-regular symmetric matrix $\begin{bmatrix}
a & e \\ e & a
\end{bmatrix}$ and its inverse
$
\begin{bmatrix}
b & f \\ f & b
\end{bmatrix}
$, satisfying
\begin{equation}
\begin{bmatrix}
a & e \\ e & a
\end{bmatrix}
\begin{bmatrix}
b & f \\ f & b
\end{bmatrix} = \begin{bmatrix}
1 & 0 \\ 0 & 1
\end{bmatrix}.
\label{eq:abef}
\end{equation}
In particular, it is required that $a$, $e$ and $a^2-e^2$ are all not equal to zero in $\mathbb{F}_q$.

\medskip

\noindent {\bf Encoding.} A data file consists of
$$B = k(d+t-k) = k(n-k) = k^2$$
source symbols. For $i=1,2,\ldots, k$, node $i$ is a systematic node and stores $k$ source symbols.
We can perform the encoding in two essentially the same ways. In the first encoding function, the first $k$ nodes store the source symbols and the last $k$ nodes store the parity-check symbols.
Let $\bx_i$ be the $k$-dimensional vector whose components are the symbols stored in node $i$. For $j=1,2,\ldots, k$, node $k+j$ is a parity-check node, and stores the $k$ components of vector
\begin{equation}
\by_j =  \sum_{\ell=1}^k \big( a \hat{\bu}_\ell \bv_j^T + e p_{\ell j}\bI_k \big) \bx_\ell  ,
\label{eq:encode1}
\end{equation}
where $\bI_k$ denotes the $k\times k$ identity matrix. We note that the matrix within the parenthesis in \eqref{eq:encode1} is the sum of a rank-1 matrix and an identity matrix.

In the second encoding function, which is the dual of the first one, nodes $k+1, k+2, \ldots, 2k$ store the source symbols and nodes $1$ to $k$ store the parity-check symbols.
Let $\by_j$ be the $k$-dimensional vector stored in node $k+j$.
For $i=1,2,\ldots, k$, node $i$ stores the vector
\begin{equation}
\bx_i =  \sum_{\ell=1}^k \big( b \hat{\bv}_\ell \bu_i^T + f q_{\ell i}\bI_k \big) \by_\ell.
\label{eq:encode2}
\end{equation}

This duality relationship is first noted in~\cite{SR11}.

\begin{prop}[\cite{SR11}]
The regenerating code defined by \eqref{eq:encode1} is the same as the one defined by \eqref{eq:encode2}.
\label{prop:dual}
\end{prop}

We will give a proof of Prop.~\ref{prop:dual} in terms of matrices. The matrix formulation is also useful in simplifying the description of the repair and decode procedure.
Let $\hat{\bU}$ (resp. $\hat{\bV}$, $\bX$ and $\bY$) be the $k\times k$ matrix whose columns are $\hat{\bu}_i$ (resp. $\hat{\bv}_i$, $\bx_i$, and $\by_i$) for $i=1,2,\ldots, k$. We have
\begin{align*}
\hat\bU &= (\bU^{-1})^T = \hat\bV (\bQ^{-1})^T, \\
\hat\bV &= (\bV^{-1})^T = \hat\bU (\bP^{-1})^T.
\end{align*}
In terms of these matrices, the first encoding function can be expressed as
\begin{align}
\bY &=a \hat{\bU} \bX^T \bV + e \bX \bP.  \label{eq:encode1a}
\end{align}
Indeed, the $j$-th column of $a \hat{\bU} \bX^T \bV + e \bX \bP$ is
\begin{align*}
&\phantom{=} a \hat{\bU} \cdot (\text{$j$-th column of $\bX^T \bV$})+ e \sum_{\ell=1}^k \bx_\ell p_{\ell j} \\
&= a \sum_{\ell=1}^k \hat{\bu}_\ell \cdot  (\bx_\ell^T \bv_j)+ e \sum_{\ell=1}^k \bx_\ell p_{\ell j} \\
&= a \sum_{\ell=1}^k \hat{\bu}_\ell \bv_j^T \bx_\ell + e \sum_{\ell=1}^k \bx_\ell p_{\ell j} \\
&= \sum_{\ell=1}^k \big( a  \hat{\bu}_\ell \bv_j^T + e p_{\ell j} \bI_k \big) \bx_\ell.
\end{align*}
Similarly, the second encoding function defined by \eqref{eq:encode2} can be expressed as
\begin{align}
\bX &= b \hat{\bV} \bY^T \bU + f \bY \bQ.  \label{eq:encode2a}
\end{align}

\begin{proof}{Proof of Prop.~\ref{prop:dual}}
Suppose that $\bY$ is given as in \eqref{eq:encode1a}. Substituting $\bY$ by   $a \hat{\bU} \bX^T \bV + e \bX \bP$ in the right-hand side of \eqref{eq:encode2a}, we get
\begin{align*}
\text{R.H.S. of \eqref{eq:encode2a}} & = b \hat{\bV} \bY^T \bU + f \bY \bQ \\
	&= b \hat{\bV} (a \hat{\bU} \bX^T \bV + e \bX \bP)^T \bU \\
	& \qquad + f (a \hat{\bU} \bX^T \bV + e \bX \bP) \bQ\\
	&= (ab +ef) \bX + (be+af) \hat{\bU} \bX^T \bU \\
	& = \bX  = \text{L.H.S. of \eqref{eq:encode2a}}.
\end{align*}
The last line follows from the facts that $ae+ef=1$ and $be+af=0$, which follow directly from \eqref{eq:abef}. Therefore,  \eqref{eq:encode2a} is implied by \eqref{eq:encode1a}.

By similar arguments, one can show that \eqref{eq:encode1a} is implied by \eqref{eq:encode2a}. Therefore,  regenerating code defined by the first encoding function in \eqref{eq:encode1} is the same as the one defined by the second encoding function in \eqref{eq:encode2}.
\end{proof}


\smallskip

\noindent {\bf Repair Procedure.}
Suppose that $t$ systematic nodes fail, for some positive integer $t \leq k$. We assume without loss of generality that the failed nodes are  nodes 1 to $t$, after some appropriate node re-labeling if necessary.


In the first phase of repair, each of the surviving nodes sends a symbol to each of the new node.  For $i=1,2,\ldots, t$, the symbol sent to node $i$ is obtained by taking the inner product of $\bu_i$  with the content of the helper node.

Consider node $i$, for some fixed index $i\in\{1,2,\ldots, t\}$.
The symbols received by node $i$ after the first phase of repair are
$$
\begin{array}{ll}
\bu_i^T \bx_m \ \ & \text{for } m = t+1,t+2,\ldots, k, \text{ and } \\
\bu_i^T \by_j \ \ & \text{for } j=1,2,\ldots, k.
\end{array}
$$
We make a change of variables and define $$\bZ := \bY \bQ.$$ For $\nu=1,2,\ldots,k$, the $\nu$-th column of $\bZ$ is
$$
\bz_\nu := \sum_{\ell=1}^k  q_{\ell \nu} \by_\ell.
$$

Because $\bQ$ is a non-singular matrix, Node $i$ can obtain the vector $(\bu_i^T  \bz_\nu)_{\nu=1,2,\ldots, k}$ from $(\bu_i^T \by_\nu)_{\nu=1,2,\ldots, k}$, and vice versa.
In terms of the new variables in $\bZ$, \eqref{eq:encode2a} becomes
\begin{equation}
\bX = b \hat\bU \bZ^T \bU + f \bZ.
 \label{eq:encode2b}
\end{equation}
The symbol sent from node $m$ to node $i$, namely $\bu_i^T \bx_m$, is the $m$-th component of vector
$$\bu_i^T \bX = \bu_i^T (b \hat{\bU} \bZ^T \bU + f \bZ),
$$
and is equal to
$$
b \bz_i^T \bu_m + f \bu_i^T \bz_m.$$

As a result, the information obtained by node $i$ after the first repair phase can be transformed to
$$
\begin{array}{ll}
\bu_i^T \bz_j & \text{for } j=1,2,\ldots, k, \text{ and }\\
b \bu_m^T \bz_i + f \bu_i^T \bz_m \ \  & \text{for } m = t+1, t+2, \ldots, k.
\end{array}
$$

In the second phase of the repair procedure, node $i$ sends the symbol $\bu_i^T \bz_{i'}$ to node $i'$, for $i,i'\in\{1,2,\ldots, t\}$, $i\neq i'$.
The total number of symbols transmitted during the first and the second part of the repair procedure is $td+t(t-1)$. The number of symbol transmissions per failed node is thus
$$
\gamma = d+t-1.
$$

\smallskip

Node $i$ wants to recover the $i$-th column of $\bX$, as expressed in \eqref{eq:encode2b}.
The $i$-th column of the first term $b \hat\bU \bZ^T \bU$ on the right-hand side is equal to the product of $b\hat \bU$ and the $i$-th column of $\bZ^T \bU$. We note that the components of the $i$-th column of $\bZ^T \bU$ are precisely $\bz_\nu^T \bu_i$, for $\nu=1,2,\ldots, k$, and are already known to node~$i$.
It remains to calculate $i$-th column of $f \bZ$, which is $f \bz_i$.

Node $i$  computes  $\bu_m^T \bz_i$ for $m=t+1,t+2,\ldots, k$ by
$$
\bu_m^T \bz_i = \frac{1}{b}[(b \bu_m^T \bz_i + f \bu_i^T \bz_m) - f \bu_i^T \bz_m].
$$
During the second phase of repair, node $i$ gets
$$
\bu_{i'}^T \bz_i, \text{for } i' \in \{1,2,\ldots, t\}\setminus\{i\}.
$$
As a result, node $i$ has a handle on $\bu_\ell^T \bz_i$ for all $\ell=1,2,\ldots,k$. Since $\bu_\ell$'s are linearly independent, node $i$ can calculate $\bz_i$ by taking the inverse of matrix $\bU$. This completes the repair procedure for node~$i$.

By dualizing the above arguments, we can collaboratively repair any $t$ parity-check node failures with optimal repair bandwidth $\gamma = d+t-1$. Note that we have not used the property that matrices $\bP$ and $\bQ$ are super-regular yet. The correctness of the repair procedure only relies on the condition that $\bP$ and $\bQ$ are non-singular.

\medskip

\noindent {\bf File Recovery.}
The reconstruction of the original file can be done in the same way as in \cite{MISER}, \cite{MISER_journal} and \cite{SR11}. We give a more concise description of the file recovery procedure below.

Suppose that a data collector connects to $k-s$ nodes among the first $k$ nodes, and $s$ nodes among the last $k$ nodes, for some integer $s$ between 0 and $k$. With suitable re-indexing, we may assume that nodes $s+1$, $s+2, \ldots, k$ are contacted by the data collector, without loss of generality. Suppose that the indices of the remaining $s$ storage nodes connected to the data collector are $j_1, j_2,\ldots, j_s$, with
$$
 k< j_1 < j_2 < \ldots < j_s \leq 2k.
$$
Thus, the data collector has access to
$$
\bx_{s+1}, \bx_{s+2},\ldots, \bx_{k}, \text{ and } \by_{j_1}, \by_{j_2}, \ldots, \by_{j_s}.
$$
The objective of the data collector is to recover vectors $\bx_1, \bx_2,\ldots, \bx_k$. Since $\bx_{s+1},\bx_{s+2},\ldots, \bx_{k}$ have been downloaded directly, we only need to reonstruct $\bx_1, \bx_2, \ldots, \bx_{s}$.

We re-write the encoding function in \eqref{eq:encode1a} as
$$ \bY = a \hat\bU \bX^T \bU \bP + e \bX \bP .
$$
The data collector only knows the columns of $\bY$ which are indexed by $j_1,j_2,\ldots, j_s$. Let
$\boldsymbol{\Upsilon}$ be the $k \times s$ submatrix of $\bY$ consisting of the columns of $\bY$ with indices $j_1, j_2,\ldots, j_s$, and let $\bpi$ be the $k\times s$ submatrix of $\bP$ consisting of columns $j_1, j_2,\ldots, j_s$. We partition matrix $\bX$ as
$$
 \bX = [\begin{array}{c|c} \bX_1 & \bX_2 \end{array}],
$$
where $\bX_1$ consists of the first $s$ columns of $\bX$, and $\bX_2$ consists of the last $k-s$ columns.

We have
\begin{equation}
\boldsymbol{\Upsilon} = a \hat\bU
\left[ \begin{array}{c}
\bX_1^T \\ \hline
\bX_2^T
\end{array}\right] \bU \bpi + e [\begin{array}{c|c} \bX_1 & \bX_2 \end{array}] \bpi .
\label{eq:MSCR_decode}
\end{equation}
Move the terms in \eqref{eq:MSCR_decode} which involve $\bX_2$ to the left, and pre-multiply by $\bU^T$. The equation in \eqref{eq:MSCR_decode} can be written as
\begin{align}
& \bU^T \boldsymbol{\Upsilon} - a
\left[ \begin{array}{c}
\mathbf{0}\\ \hline
(\bU^T \bX_2)^T
\end{array}\right]  \bpi - e [\begin{array}{c|c} \mathbf{0} & \bU^T\bX_2 \end{array}] \bpi \notag \\
& = a
\left[ \begin{array}{c}
(\bU^T \bX_1)^T \\ \hline
\mathbf{0}
\end{array}\right]  \bpi + e [\begin{array}{c|c} \bU^T \bX_1 & \mathbf{0} \end{array}] \bpi .
\label{eq:upsilon}
\end{align}
The quantities on the left of \eqref{eq:upsilon} are readily computable by the data collector.

We illustrate how to obtain $\bU^T \bX_1$ below.
Partition matrix $\bpi$ and $\bU^T \bX_1$ into
$$
\bpi = \begin{bmatrix}
\bpi_1 \\ \hline
\bpi_2
\end{bmatrix} \text{ and }
\bU^T \bX_1 =
\begin{bmatrix}
\bW_1 \\ \hline
\bW_2
\end{bmatrix},
$$
where $\bpi_1$ and $\bW_1$ are square matrices of size $s\times s$, and $\bpi_2$ and $\bW_2$ have size $(k-s)\times s$.
The right-hand side of \eqref{eq:upsilon} can be simplified to
$$
a \begin{bmatrix}
\bW_1^T \bpi_1 + \bW_2^T  \bpi_2 \\ \hline
\mathbf{0}
\end{bmatrix} +
e \begin{bmatrix}
\bW_1 \bpi_1 \\ \hline
\bW_2 \bpi_1
\end{bmatrix}.
$$
Since $\bP$ is super-regular, $\bpi_1$ is nonsingular. From the last $k-s$ rows of the matrices on both sides of \eqref{eq:upsilon}, we can solve for the entries in $\bW_2$. It remains to solve for the entries in $\bW_1$.

As the entries in $\bW_2$ are known as this point, we can subtract $a \bW_2^T \bpi_2$ from the first $s$ rows of \eqref{eq:upsilon}. We thus know the value of
$$
a \bW_1^T \bpi_1 + e \bW_1 \bpi_1.
$$
As $\bpi_1$ is nonsingular, we can post-multiply by the inverse of $\bpi_1$ and compute the $s\times s$ matrix
$$
a \bW_1^T  + e \bW_1 .
$$
The diagonal entries are $(a+e) w_{\ell \ell}$, for $\ell=1,2,\ldots ,s$.
Because $a^2-e^2=(a+e)(a-e)$ is not equal to 0 by \eqref{eq:abef}, we can divide by  $a+e$ and obtain $w_{\ell \ell}$. The non-diagonal entries can be calculated in pairs. For $i\neq j$, we solve for $w_{ij}$ and $w_{ji}$ from
$$
\begin{bmatrix}
a & e \\ e & a
\end{bmatrix}
\begin{bmatrix}
w_{ij} \\ w_{ji}
\end{bmatrix}.
$$
The above $2\times 2$ is nonsingular by \eqref{eq:abef}. Putting matrices $\bW_1$ and $\bW_2$ together, we get  $\bU^T \bX_1$. Since $\bU$ is invertible, we can solve for $\bX_1$, which consists of the vectors stored in the first $s$ storage nodes. This completes the file recovery procedure.

\medskip

\begin{table*}[t]
\caption{Encoding of a rate-$1/2$ MSCR code for eight storage nodes. The symbols in the first four columns are the source symbols in nodes 1 to 4. The symbols in the last four columns are the parity-check symbols in nodes 5 to 8.}
\label{fig:MSCR}
\begin{center}
\begin{tabular}{|c|c|c|c|c|c|c|c|} \hline
N1 & N2 & N3 & N4  & N5 & N6 &N7 & N8    \\ \hline \hline
$x_{11}$ & $x_{12} $& $x_{31}$ & $x_{41}$ & $(2x_{1\ell}+x_{\ell1})_{\ell=1}^4\cdot \bp_1$ & $(2x_{1\ell}+x_{\ell1})_{\ell=1}^4\cdot \bp_2$ & $(2x_{1\ell}+x_{\ell1})_{\ell=1}^4\cdot \bp_3$ & $(2x_{1\ell}+x_{\ell1})_{\ell=1}^4\cdot \bp_4$  \\
$x_{12}$ & $x_{22}$ & $x_{32}$ & $x_{42}$ & $(2x_{2\ell}+x_{\ell2})_{\ell=1}^4\cdot \bp_1$ & $(2x_{2\ell}+x_{\ell2})_{\ell=1}^4\cdot \bp_2$ & $(2x_{2\ell}+x_{\ell2})_{\ell=1}^4\cdot \bp_3$ & $(2x_{2\ell}+x_{\ell2})_{\ell=1}^4\cdot \bp_4$  \\
$x_{13}$ & $x_{23}$& $x_{33}$ & $x_{43}$ & $(2x_{3\ell}+x_{\ell3})_{\ell=1}^4\cdot \bp_1$ & $(2x_{3\ell}+x_{\ell3})_{\ell=1}^4\cdot \bp_2$ & $(2x_{3\ell}+x_{\ell3})_{\ell=1}^4\cdot \bp_3$ & $(2x_{3\ell}+x_{\ell3})_{\ell=1}^4\cdot \bp_4$     \\
$x_{14}$ & $x_{24}$ & $x_{34}$ & $x_{44}$ & $(2x_{4\ell}+x_{\ell4})_{\ell=1}^4\cdot \bp_1$ & $(2x_{4\ell}+x_{\ell4})_{\ell=1}^4\cdot \bp_2$ & $(2x_{4\ell}+x_{\ell4})_{\ell=1}^4\cdot \bp_3$ & $(2x_{4\ell}+x_{\ell4})_{\ell=1}^4\cdot \bp_4$   \\ \hline
\end{tabular}
\end{center}

\end{table*}

{\bf Example.} Consider an example for $k=4$. There are eight storage nodes in the distributed storage system. Nodes 1 to 4 are the systematic nodes, while nodes 5 to 8 are the parity-check nodes. The data file contains $B=k^2=16$ symbols in a finite field. For $i=1,2,3,4$, we let the symbols stored in node $i$ be denoted by $x_{i1}$, $x_{i2}$, $x_{i3}$ and $x_{i4}$ (see Table~\ref{fig:MSCR}). In this example, we pick a finite field of size 11 as the alphabet. All arithmetic is performed modulo 11.

We let $\bP$ be the following $4\times 4$ Cauchy matrix
$$
\bP=
\begin{bmatrix}
  \frac{1}{a_1-b_1} &
  \frac{1}{a_1-b_2} &
  \frac{1}{a_1-b_3} &
  \frac{1}{a_1-b_4} \\
  \frac{1}{a_2-b_1} &
  \frac{1}{a_2-b_2} &
  \frac{1}{a_2-b_3} &
  \frac{1}{a_2-b_4} \\
  \frac{1}{a_3-b_1} &
  \frac{1}{a_3-b_2} &
  \frac{1}{a_3-b_3} &
  \frac{1}{a_3-b_4} \\
  \frac{1}{a_4-b_1} &
  \frac{1}{a_4-b_2} &
  \frac{1}{a_4-b_3} &
  \frac{1}{a_4-b_4}
\end{bmatrix} =
\begin{bmatrix}
1 & 4 & 9 & 8 \\
10 & 1 & 4 & 9 \\
7 &   10 &   1 &   4 \\
2 &   7 &   10 &   1
\end{bmatrix}
$$
where $a_1=2$, $a_2=4$, $a_3=6$, $a_4=8$, $b_1=1$, $b_2=3$, $b_3=5$ and $b_4=7$ are distinct elements in $\mathbb{F}_{11}$. Matrices $\bU$ and $\hat\bU$ are set to the $4\times 4$ identity matrix. With this choice of matrix $\bU$, the matrix $\bV$ is equal to $\bU\bP = \bP$.
Let $a=2$ and $e=1$, such that the matrix
$$
 \begin{bmatrix} a & e \\ e & a \end{bmatrix} =
 \begin{bmatrix} 2 & 1 \\ 1 & 2 \end{bmatrix}
$$
is super-regular. For $j=1,2,3,4$, the symbols stored in the $j$-th parity-check node are the entries in the $j$-th column of the following matrix
$$
\bY = \left( 2\begin{bmatrix}
x_{11}&x_{12}&x_{13}&x_{14} \\
x_{21}&x_{22}&x_{23}&x_{24} \\
x_{31}&x_{32}&x_{33}&x_{34} \\
x_{41}&x_{42}&x_{43}&x_{44}
\end{bmatrix}
+
\begin{bmatrix}
x_{11}&x_{21}&x_{31}&x_{41} \\
x_{12}&x_{22}&x_{32}&x_{42} \\
x_{13}&x_{23}&x_{33}&x_{43} \\
x_{14}&x_{24}&x_{34}&x_{44}
\end{bmatrix} \right) \bP.
$$
For $j=1,2,3,4$, we denote the $j$-th column of the Cauchy matrix $\bP$ by $\bp_j$, and let $(2x_{j\ell}+ x_{\ell j})_{\ell=1}^4$ be the vector
\begin{equation}
(2x_{j1}+x_{1j}, 2x_{j2}+x_{2j}, 2x_{j3}+x_{3j}, 2x_{j4}+x_{4j}).
\label{eq:v}
\end{equation}
The content of the parity-check nodes are illustrated in the last four columns in Table~\ref{fig:MSCR}.

We illustrate how to repair multiple systematic node failures collaboratively.
Suppose that nodes 1, 2 and 3 fail.
In the first phase of the repair process, each of the remaining nodes, namely nodes 4 to 8, sends three symbols to each new node. In this example, each helper node can simply read out three symbols and send them to the new nodes. More specifically, the symbols in the first (resp. second and third) row in columns 4 to 8 in Table~\ref{fig:MSCR} are sent to node 1 (resp. 2 and 3). Hence, for $i=1,2,3$, new node $i$ receives the following
five finite field symbols
\begin{gather*}
 x_{4i}, \ (2x_{i\ell}+x_{\ell i})_{\ell=1}^4\cdot \bp_1,\ (2x_{i\ell}+x_{\ell i})_{\ell=1}^4\cdot \bp_2, \\
(2x_{i \ell}+x_{\ell i})_{\ell=1}^4\cdot \bp_3,\ (2x_{i\ell}+x_{\ell i})_{\ell=1}^4\cdot \bp_4
\end{gather*}
in the first phase. Since $\bP$ is a nonsingular matrix, new node $i$ can obtain the vector $(2x_{i\ell}+x_{\ell i})_{\ell=1}^4$. We list the symbols which can be computed by the new nodes as follows,

\medskip

\noindent Node 1:\  $x_{41}$, $3x_{11}$, $2x_{12}+x_{21}$, $2x_{13}+x_{31}$, $2x_{14}+x_{41}$.

\smallskip

\noindent Node 2:\  $x_{42}$, $2x_{21}+x_{12}$, $3x_{22}$, $2x_{23}+x_{32}$, $2x_{24}+x_{42}$.

\smallskip

\noindent Node 3:\  $x_{43}$, $2x_{31}+x_{13}$, $2x_{32}+x_{23}$, $3x_{33}$, $2x_{34}+x_{43}$.

\smallskip

At the end of the first phase, new node $i$ can calculate the $i$th symbol $x_{ii}$ and the last symbol $x_{i4}$ by
\begin{align*}
x_{ii} &= \frac{1}{3} (3x_{ii}) \\
 x_{i4} &= \frac{1}{2}[(2x_{i4}+x_{4i}) - x_{4i}].
\end{align*}

The operations in the second phase of the repair are:

\noindent  Node 1 and node 2 exchange the symbols $2x_{12}+x_{21}$ and $2x_{21}+x_{12}$;

\noindent  Node 1 and node 3 exchange the symbols $2x_{13}+x_{31}$ and $2x_{31}+x_{13}$;

\noindent  Node 2 and node 3 exchange the symbols $2x_{23}+x_{32}$ and $2x_{32}+x_{23}$.

Now, nodes 1 can decode the symbols $x_{12}$ and $x_{13}$ from
$$
\begin{bmatrix} 2 & 1 \\ 1 & 2 \end{bmatrix}
\begin{bmatrix} x_{12} \\ x_{21} \end{bmatrix} \text{ and }
\begin{bmatrix} 3 & 1 \\ 1 & 3 \end{bmatrix}
\begin{bmatrix} x_{13} \\ x_{31} \end{bmatrix}.
$$
Similarly, nodes 2 and 3 can decode the remaining sources symbols.

We transmitted 21 symbols during the whole repair procedure. Hence, 7 symbol transmissions are required per new node. It matches the lower bound on repair bandwidth per new node
$$\gamma_\text{MSCR}= d+t-1 = 5+3-1 = 7.
$$

{\it Remark:} In the previous example, can see the use of {\bf interference alignment} as follows.  After the first phase of repair, the first new node has symbols $x_{41}$, $3x_{11}$, $2x_{12}+x_{21}$, $2x_{13}+x_{31}$, $2x_{14}+x_{41}$, but the first new node is only interested in decoding symbols $x_{11}$, $x_{12}$, $x_{13}$ and $x_{14}$. The symbols $x_{21}$, $x_{31}$, and $x_{41}$   can be regarded as ``interference'' with respect to the first new node. The interference occupy three degrees of freedom and is resolved in the second phase of the repair.

\medskip


Suppose that a data collector wants to recover the original file by downloading the symbols stored in nodes 3, 4, 5 and 6. The symbols stored in nodes 3 and 4 are uncoded symbols, and hence can be read off directly. The data collector needs to decode $x_{11}$, $x_{12}$, $x_{13}$, $x_{14}$, $x_{21}$, $x_{22}$, $x_{23}$ and $x_{24}$ from symbols in node 5,
\begin{gather}
(3x_{11}, \ 2x_{12}+x_{21},\  2x_{13}+\underline{x_{31}}, \ 2x_{14}+\underline{x_{41}}) \cdot \bp_1 \label{eq:decode_a} \\
(2x_{21}+x_{12}, \ 3x_{22},\  2x_{23}+\underline{x_{32}}, \ 2x_{24}+\underline{x_{41}}) \cdot \bp_1 \label{eq:decode_b} \\
(2\underline{x_{31}}+x_{13}, \ 2\underline{x_{32}}+x_{23},\  3\underline{x_{33}}, \ 2\underline{x_{34}}+\underline{x_{43}}) \cdot \bp_1 \label{eq:decode_c} \\
(2\underline{x_{41}}+x_{14}, \ 2\underline{x_{42}}+x_{24},\  2\underline{x_{43}}+\underline{x_{34}}, \ 3\underline{x_{44}}) \cdot \bp_1  \label{eq:decode_d}
\end{gather}
and the symbols in node 6,
\begin{gather}
(3x_{11}, \ 2x_{12}+x_{21},\  2x_{13}+\underline{x_{31}}, \ 2x_{14}+\underline{x_{41}}) \cdot \bp_2 \label{eq:decode_e} \\
(2x_{21}+x_{12}, \ 3x_{22},\  2x_{23}+\underline{x_{32}}, \ 2x_{24}+\underline{x_{41}}) \cdot \bp_2 \label{eq:decode_f}\\
(2\underline{x_{31}}+x_{13}, \ 2\underline{x_{32}}+x_{23},\  3\underline{x_{33}}, \ 2\underline{x_{34}}+\underline{x_{43}}) \cdot \bp_2 \label{eq:decode_g}\\
(2\underline{x_{41}}+x_{14}, \ 2\underline{x_{42}}+x_{24},\  2\underline{x_{43}}+\underline{x_{34}}, \ 3\underline{x_{44}}) \cdot \bp_2 . \label{eq:decode_h}
\end{gather}
The underlined symbols are readily obtained from nodes 3 and 4.

From the two finite field symbols in \eqref{eq:decode_d} and \eqref{eq:decode_h}, after subtracting off the known quantities, we can decode symbol $x_{14}$ and $x_{24}$ from
$$
\begin{bmatrix} x_{14} &\  x_{24} \end{bmatrix} \cdot \bpi_1,
$$
where $$\bpi_1 = \begin{bmatrix} p_{11} & p_{12} \\ p_{21} & p_{22}\end{bmatrix}
= \begin{bmatrix} 1&4 \\ 10 &1 \end{bmatrix} $$
 is the $2\times 2$ submatrix on the top left corner of $\bP$.
Likewise, from \eqref{eq:decode_c} and \eqref{eq:decode_g}, we can obtain $x_{13}$ and $x_{23}$ by solving a $2\times 2$ system of linear equations.

We can put the four finite field symbols in\eqref{eq:decode_a}, \eqref{eq:decode_b}, \eqref{eq:decode_e} and \eqref{eq:decode_f} together and form a $2 \times 2$ matrix
$$
\begin{bmatrix}
3 x_{11} & 2 x_{12}+x_{21} \\
2 x_{21}+x_{12} & 3 x_{22}
\end{bmatrix} \bpi_1.
$$
Using the property that $\bpi_1$ is non-singular again, we can solve for the matrix
$$
\begin{bmatrix}
3 x_{11} & 2 x_{12}+x_{21} \\
2 x_{21}+x_{12} & 3 x_{22}
\end{bmatrix},
$$
from which we can decode $x_{11}$, $x_{22}$, $x_{12}$ and $x_{21}$.

\section{Concluding Remarks}
\label{sec:conclusion}

In this paper we review two constructions of cooperative regenerating codes, one for the MSCR point and one for the MBCR point. We show that with the same coding structure as in the MISER code, we can cooperatively repair any number of  systematic node failures and any number of parity-check node failures. As a matter of fact, we can also repair any pair of systematic node and a parity-check node. However, we need to work over a larger finite field and the super-regular matrix $\mathbf{P}$ should satisfy some extra conditions, in order to repair any two node failures. The detail can be found in~\cite{ChenShum}.

Security aspects of cooperative regenerating codes are investigated in~\cite{OD11} to
\nocite{KRV14} \cite{HPX16}. There are basically two types of adversarial storage nodes. Adversaries of the first type are passive eavesdroppers, who want to obtain some information about the data file. Under the assumption that the number of storage nodes accessed by an eavesdropper is no more than a certain number, the secrecy capacity and the related code constructions are studied in~\cite{KRV14} and \cite{HPX16}. Adversaries of the second type, called Byzantine adversaries, are malicious and try to corrupt the distributed storage system. They conform to the protocol but may send out erroneous packets during a repair procedure. It is shown in~\cite{OD11} that distributed storage system with cooperative repair is more susceptible to this kind of pollution attack, because of the large number of data exchanges in the second phase of the repair. One way to alleviate the potential damage incurred by a Byzantine adversary is to allow multiple levels of cooperation. To this end,
a partially cooperative repair model, in which a new node communicates only with a fraction of all new nodes, is proposed in \cite{LO14a}. A code construction based on subspace codes is given in \cite{LO14b}.

Local repairable code (LRC) is another class of erasure-correcting codes of practical interests. In LRC, the focus is not on the repair bandwidth, but on the number of nodes contacted by a new node.
A code is said to have {\em locality} $r$ if each symbol in a codeword is a function of at most $r$ other symbols. In contrast to regenerating code, it is only required that, for each symbol, there exists a particular set of $r$ nodes from which we can repair the symbol.
A fundamental bound on the minimum distance of a code with locality constraint was obtained by Gopalan {\em et al.} in \cite{Gopalan12}. The problem of repairing multiple symbol errors locally, called
{\em cooperative local repair}, is studied in \cite{cooperative_local_repair} and \cite{LGSF16}.

In addition to locality, disk I/O cost is another important factor. The speed of reading bits from hard disks may be a bottleneck of the repair time. The number of bits that must be accessed by a helper node is obviously lower  bounded by the number of bits transmitted to the new nodes. A regenerating code with the property that the number of bits accessed for the purpose of repaired is exactly equal to the number of bits transmitted is called a {\em repair-by-transfer} or {\em help-by-transfer} code (see e.g. \cite{SAK15} or \cite{SRKR12}). We note that the example in Section~\ref{sec:MBCR} is indeed a repair-by-transfer MBCR code, even though the repair is for systematic nodes only. It is proved in \cite{WangZhang} that repair-by-transfer MBCR code does not exists when if any $t\geq 2$ failed nodes could be repaired by any $d\geq 2$ helper nodes. The example in Section~\ref{sec:MBCR} does not violate the impossibility result in \cite{WangZhang}. Nevertheless, it is interesting to see whether repair-by-transfer is possible for other code parameters.

Instead of designing new codes, devising efficient algorithms which can repair existing storage codes is also of practical interests. Fast repair method for the traditional Reed-Solomon code can be found in~\cite{RepairingRS}. Recovery algorithm for array codes, such as Row Diagonal Pairty (RDP) and X-code, are given in \cite{RDP_repair} and \cite{Xcode_repair}, respectively. Some special results for repairing concurrent failures in RDP code are reported in \cite{RDP_concurrent_repair}. It is interesting to see whether we can devise cooperative repair algorithm for Reed-Solomon codes and other array codes.


\end{document}